\documentstyle{article}
       \renewenvironment{abstract}{\section*{Abstract}\small}{}
       \newtheorem{definition}{Definition}

       \newtheorem{theorem}[definition]{Theorem}
       
       \newtheorem{lemma}[definition]{Lemma}

       \makeatletter
       \renewcommand{\@begintheorem}[2]{ 
         \trivlist\item[\hskip\labelsep{\bf #1\ #2}]}
       \renewcommand{\@opargbegintheorem}[3]{\trivlist
         \item[\hskip \labelsep{\bf #1\ #2\ (#3)}]}
       \makeatother
       \newtheorem{proof}{Proof}
       
\begin{document}

\begin{center}
{\large\bf Uniform Provability in Classical Logic}\\[7pt]
{\it Gopalan Nadathur}\\[5pt]
Department of Computer Science\\
University of Chicago\\
Ryerson Hall\\
1100 E 58th Street\\
Chicago, IL 60637\\[5pt]
Phone Number: (773)-702-3497\\[5pt]
Fax Number: (773)-702-8487\\[5pt]
Email: \verb+gopalan@cs.uchicago.edu+
\end{center}
\bigskip\bigskip


\newdimen\linewidth
\linewidth=\hsize

\def\endfake{\bigskip\bigskip}
\def\mud#1{\hfil $\displaystyle{#1}$\hfil}
\def\rig#1{\hfil $\displaystyle{#1}$}

\def\vcaligntext{\noindent\vcalignhbox}
\def\vcalignhbox#1{\hbox{\valign{\vfil\hbox{##}\vfil\cr #1}}}
\def\vcalign#1{$$\vcalignhbox{#1}$$}
\def\vcalignn#1{\noindent{\small\hbox to \linewidth{%
  \hss\valign{\vfil\hbox{##}\vfil\cr #1}\hss}}}

\def\lowerhalf#1{\hbox{\raise -0.5\baselineskip\hbox{#1}}}


\def\lnot{\mathord{\sim}}
\def\lforall{\forall}
\def\lexists{\exists}
\def\limplies{\mathbin{\supset}}
\def\lelbow{\neg}
\def\larrow{\mathbin{\rightarrow}}
\def\lbottom{\mathord{\perp}}

\def\cut{\mathord{Cut}}
\def\ori{\mathord{\lor{\rm I}}}
\def\andi{\mathord{\land{\rm I}}}
\def\existsi{\mathord{\lexists{\rm I}}}

\def\goesto{\hbox to 0pt{\hss $\Rightarrow$ \hss}}


\newbox\tempa
\newbox\tempb
\newdimen\tempc
\newbox\tempd


\def\inruleanhelp#1#2#3{\setbox\tempa=\hbox{$\displaystyle{\mathstrut #2}$}%
                        \setbox\tempb=\vbox{\halign{##\cr
        \mud{#1}\cr
        \noalign{\vskip \the\lineskip}%
        \noalign{\vskip 1pt}%
        \noalign{\hrule height 0pt}%
        \rig{\vbox to 0pt{\vss\hbox to 0pt{${\ #3}$\hss}\vss}}\cr
        \noalign{\hrule}%
        \noalign{\vskip \the\lineskip}%
        \mud{\copy\tempa}\cr}}%
                      \tempc=\wd\tempb
                      \advance\tempc by \wd\tempa
                      \divide\tempc by 2 }

\def\inrulemhelp#1#2#3{\setbox\tempa=\hbox{$\displaystyle{\mathstrut #2}$}%
                        \setbox\tempb=\vbox{\halign{##\cr
        \mud{#1}\cr
        \noalign{\vskip\the\lineskip}%
        \noalign{\hrule}%
        \noalign{\vskip\the\lineskip}%
        \noalign{\vskip 1pt}%
        \noalign{\hrule height 0pt}%
        \rig{\vbox to 0pt{\vss\hbox to 0pt{${\ #3}$\hss}\vss}}\cr
        \noalign{\hrule}%
        \noalign{\vskip\the\lineskip}%
        \mud{\copy\tempa}\cr}}%
                      \tempc=\wd\tempb
                      \advance\tempc by \wd\tempa
                      \divide\tempc by 2 }


\def\inrulemchelp#1#2#3{\setbox\tempa=\hbox{$\displaystyle{\mathstrut #2}$}%
                        \setbox\tempd=\hbox{$\ #3$}%
                        \setbox\tempb=\vbox{\halign{##\cr
        \mud{#1}\cr
        \noalign{\vskip\the\lineskip}%
        \noalign{\vskip 1pt}%
        \noalign{\hrule}%
        \noalign{\vskip\the\lineskip}%
        \noalign{\hrule height 0pt}%
        \rig{\vbox to 0pt{\vss\hbox to 0pt{${\ #3}$\hss}\vss}}\cr
        \noalign{\hrule}%
        \noalign{\vskip\the\lineskip}%
        \mud{\copy\tempa}\cr}}%
                      \tempc=\wd\tempb
                      \advance\tempc by \wd\tempa
                      \divide\tempc by 2 }


\def\inruleanchelp#1#2#3{\setbox\tempa=\hbox{$\displaystyle{\mathstrut #2}$}%
                        \setbox\tempd=\hbox{$\ #3$}%
                        \setbox\tempb=\vbox{\halign{##\cr
        \mud{#1}\cr
        \noalign{\vskip\the\lineskip}%
        \noalign{\vskip 1pt}%
        \noalign{\hrule height 0pt}%
        \rig{\vbox to 0pt{\vss\hbox to 0pt{\copy\tempd \hss}\vss}}\cr
        \noalign{\hrule}%
        \noalign{\vskip\the\lineskip}%
        \mud{\copy\tempa}\cr}}%
                      \tempc=\wd\tempb
                      \advance\tempc by \wd\tempa
                      \divide\tempc by 2 }



\def\inrulean#1#2#3{{\inruleanhelp{#1}{#2}{#3}%
                     \hbox to \wd\tempa{\hss \box\tempb \hss}}}

\def\inruleanc#1#2#3{{\inruleanchelp{#1}{#2}{#3}%
                \lowerhalf{\box\tempb\hskip\wd\tempd}}}

\def\linrulean#1#2#3{{\inruleanhelp{#1}{#2}{#3}%
                      \hbox to \tempc{\hss \box\tempb}}}

\def\rinrulean#1#2#3{{\inruleanhelp{#1}{#2}{#3}%
                      \hbox to \tempc{\box\tempb \hss}}}

\def\ginrulean#1#2#3{{\inruleanhelp{#1}{#2}{#3}%
                     \hbox to .5\linewidth{\hfil \box\tempb \hfil}}}

\def\ginruleanl#1#2#3{{\inruleanchelp{#1}{#2}{#3}%
                       \hbox to .5\linewidth{\hfil
                         \box\tempb\hskip\wd\tempd \hfil}}} 


\def\inrulem#1#2#3{{\inrulemhelp{#1}{#2}{#3}%
                    \hbox to \wd\tempa{\hss \box\tempb \hss}}}

\def\inrulemc#1#2#3{{\inrulemchelp{#1}{#2}{#3}%
                     \lowerhalf{\box\tempb\hskip\wd\tempd}}}

\def\linrulem#1#2#3{{\inrulemhelp{#1}{#2}{#3}%
                      \hbox to \tempc{\hss \box\tempb}}}

\def\rinrulem#1#2#3{{\inrulemhelp{#1}{#2}{#3}%
                      \hbox to \tempc{\box\tempb \hss}}}

\def\ginrulem#1#2#3{{\inrulemhelp{#1}{#2}{#3}%
                     \hbox to .5\linewidth{\hfil \box\tempb \hfil}}}

\def\ginruleml#1#2#3{{\inrulemchelp{#1}{#2}{#3}%
                     \hbox to .5\linewidth{\hfil
                       \box\tempb \hskip\wd\tempd \hfil}}}


\def\inrulebn#1#2#3#4{\inrulean{#1\qquad\qquad #2}{#3}{#4}}
\def\inrulebnc#1#2#3#4{\inruleanc{#1\qquad\qquad #2}{#3}{#4}}
\def\linrulebn#1#2#3#4{\linrulean{#1\qquad\qquad #2}{#3}{#4}}
\def\ginrulebn#1#2#3#4{\ginrulean{#1\qquad\qquad #2}{#3}{#4}}
\def\ginrulebnl#1#2#3#4{\ginruleanl{#1\qquad\qquad #2}{#3}{#4}}
\def\rinrulebn#1#2#3#4{\rinrulean{#1\qquad\qquad #2}{#3}{#4}}


\newbox\gappremises

\def\gap#1{\gapbuild#1{}\inrulegap{\box\gappremises}}

\def\gapbuild#1{\if *#1*\let\next=\relax\else
        \setbox\gappremises=\hbox{\ifvoid\gappremises\else
                                         \unhbox\gappremises \qquad\fi
                                  \mud{#1}}%
        \let\next=\gapbuild\fi\next}

\def\gaphelp#1#2#3{\setbox\tempa=\hbox{$\displaystyle{\mathstrut #2}$}%
                   \setbox\tempb=\vbox{\lineskip=2pt%
 \halign{##\cr
        \mud{#1}\cr
        \noalign{\hrule height 0pt}%
        \mud{#3}\cr
        \noalign{\hrule height 0pt}%
        \noalign{\vskip\the\lineskip}%
        \mud{\copy\tempa}\cr}}%
                      \tempc=\wd\tempb
                      \advance\tempc by \wd\tempa
                      \divide\tempc by 2 }

\def\gapchelp#1#2#3{\setbox\tempa=\hbox{$\displaystyle{\mathstrut #2}$}%
                   \setbox\tempb=\hbox{$\vcenter{\lineskip=2pt%
 \halign{##\cr
        \mud{#1}\cr
        \noalign{\hrule height 0pt}%
        \mud{#3}\cr
        \noalign{\hrule height 0pt}%
        \noalign{\vskip\the\lineskip}%
        \mud{\copy\tempa}\cr}}$}%
                      \tempc=\wd\tempb
                      \advance\tempc by \wd\tempa
                      \divide\tempc by 2 }


\def\inrulegap#1#2#3{{\gaphelp{#1}{#2}{#3}%
                      \hbox to \wd\tempa{\hss \box\tempb \hss}}}

\def\ginrulegap#1#2#3{{\gaphelp{#1}{#2}{#3}%
                       \hbox to 0.5\linewidth{\hfil \box\tempb \hfil}}}

\def\inrulegapc#1#2#3{{\gaphelp{#1}{#2}{#3}%
                     \lowerhalf{\box\tempb}}}


\def\inrulewjanhelp#1#2{\setbox\tempa=\hbox{$\displaystyle{\mathstrut #2}$}%
                        \setbox\tempb=\vbox{\halign{##\cr
        \mud{#1}\cr
        \noalign{\vskip\the\lineskip}%
        \noalign{\vskip 1pt}%
        \noalign{\hrule height 0pt}%
        \noalign{\hrule}%
        \noalign{\vskip\the\lineskip}%
        \mud{\copy\tempa}\cr}}%
                      \tempc=\wd\tempb
                      \advance\tempc by \wd\tempa
                      \divide\tempc by 2 }


\def\inrulewjan#1#2{{\inrulewjanhelp{#1}{#2}%
                     \hbox to \wd\tempa{\hss \box\tempb \hss}}}

\newcommand{\sep}{\;\vert\;}

\newcommand{\ra}{\rightarrow}
\newcommand{\app}{\ }
\newcommand{\appt}{\ }
\newcommand{\tup}[1]{\langle\nobreak#1\nobreak\rangle}

\newcommand{\oprove}{\vdash\kern-.6em\lower.5ex\hbox{$\scriptstyle O$}\,}
\newcommand{\cprove}{\vdash\kern-.6em\lower.5ex\hbox{$\scriptstyle C$}\,}
\newcommand{\iprove}{\vdash\kern-.6em\lower.5ex\hbox{$\scriptstyle I$}\,}

\newcommand{\all}{\forall}
\newcommand{\some}{\exists}
\newcommand{\lambdax}[1]{\lambda #1\,}
\newcommand{\somex}[1]{\some#1\,}
\newcommand\allx[1]{\all#1\,}

\newcommand{\subs}[3]{[#1/#2]#3}
\newcommand{\rep}[3]{S^{#2}_{#1}{#3}}
\newcommand{\ie}{{\em i.e.}}
\newcommand{\eg}{{\em e.g.}}

\newcommand{\mthcontr}{\mbox{\rm contr-R}}
\newcommand{\mthcontl}{\mbox{\rm contr-L}}
\newcommand{\contl}{$\mbox{\rm contr-L}$}
\newcommand{\contr}{$\mbox{\rm contr-R}$}
\newcommand{\botr}{$\bot$-R} 
\newcommand{\mthbotr}{\bot\mbox{\rm -R}}
\newcommand{\andl}{$\land$-L} 
\newcommand{\mthandl}{\land\mbox{\rm -L}} 
\newcommand{\andr}{$\land$-R}
\newcommand{\mthandr}{\land\mbox{\rm -R}} 
\newcommand{\orl}{$\lor$-L}
\newcommand{\mthorl}{\lor\mbox{\rm -L}} 
\newcommand{\orlg}{$\lor$-L$_G$}
\newcommand{\mthorlg}{\lor\mbox{\rm -L}_G} 
\newcommand{\orr}{$\lor$-R}
\newcommand{\mthorr}{\lor\mbox{\rm -R}}
\newcommand{\impl}{$\supset$-L}
\newcommand{\mthimpl}{\supset\!\mbox{\rm -L}}
\newcommand{\impr}{$\supset$-R}
\newcommand{\mthimpr}{\supset\!\mbox{\rm -R}}
\newcommand{\negl}{$\neg$-L} 
\newcommand{\alll}{$\forall$-L} 
\newcommand{\mthalll}{\forall\mbox{\rm -L}} 
\newcommand{\allr}{$\forall$-R}
\newcommand{\mthallr}{\forall\mbox{\rm -R}}
\newcommand{\somel}{$\exists$-L}
\newcommand{\mthsomel}{\exists\mbox{\rm -L}}
\newcommand{\somer}{$\exists$-R}
\newcommand{\mthsomer}{\exists\mbox{\rm -R}}
\newcommand{\resg}{\mbox{\rm res}$_G$}
\newcommand{\mthresg}{\mbox{\rm res}_G} 

\newcommand{\sequent}[2]{\hbox{{$#1\ 
      \longrightarrow\ #2$}}}

\newcommand{\Ibf}{{\bf I}}
\newcommand{\Ibfgprime}{{\bf I$^\prime_G$}}
\newcommand{\Ibfg}{{\bf I}$_G$}
\newcommand{\Obfg}{{\bf O}$_G$}
\newcommand{\Cbf}{{\bf C}}

\newsavebox{\partfig}

\newsavebox{\pfig}

\begin{abstract}
\noindent 
Uniform proofs are sequent calculus proofs with the
following characteristic: the last step in the derivation of a complex
formula at any stage in the proof is always the introduction of the
top-level logical symbol of that formula. 
We investigate the relevance of this uniform proof notion to
structuring proof search in classical logic. 
A logical language in whose context provability is equivalent to
uniform provability admits of a goal-directed proof procedure that
interprets logical symbols as search directives whose meanings are
given by the corresponding inference rules. 
While this uniform provability property does not hold directly of
classical logic, we show that it holds of a fragment of it that only
excludes essentially positive occurrences of universal quantifiers
under a modest,  
sound, modification to the set of assumptions: the addition to
them of the negation of the formula being proved. We further note that
all uses of the added formula can be factored into certain derived
rules. The resulting proof system and the uniform provability property
that holds of it are used to outline a proof procedure for
classical logic. An interesting aspect of this proof procedure is that
it incorporates within it previously proposed mechanisms for dealing
with disjunctive information in assumptions and for handling
hypotheticals. Our analysis sheds light on the relationship between
these mechanisms and the notion of uniform proofs.
\end{abstract}

\bigskip



\section{Introduction}\label{sec:intro}

Uniform proofs as identified in \cite{MNPS91} capture a
goal-directedness in proof search. In essence, a uniform 
proof is a sequent calculus proof that is found by constructing, at
each stage, a proof for a {\it single} ``goal'' formula from a
collection of assumptions. Further, if the goal formula is
non-atomic, then the search for a uniform proof for it may proceed
by first simplifying the formula in accordance with the inference
rule pertaining to its top-level logical symbol. One reason for
interest in this category of proofs is that it provides a framework
for interpreting the logical symbols in the formulas being proved as
primitives for directing search and the inference rules pertaining to
these symbols as specifications of their search semantics. This
viewpoint is exploited in \cite{MNPS91} in describing a
proof-theoretic foundation for logic programming. In particular,
classes of formulas and proof relations are thought to constitute a
satisfactory basis for logic programming just in case provability in
their context is equivalent to the existence of a uniform proof. The
virtue of this ``uniform provability property'' is that it permits a
duality between a declarative and a search-related reading for logical
symbols that appears to be central to a programming use of logic. This
criterion for logic programming has turned out to be of actual
practical interest: it is satisfied by the logic of Horn clauses that 
underlies Prolog and has also been instrumental in the discovery of
rich and useful but yet logically principled extensions to this
language \cite{HodMil94,Mil94lics,MNPS91,NM94,Pfe89}. 

Our interest in this paper is in a different, but related, utility for
uniform proofs, namely, as a device for structuring the search for
proofs of formulas. 
A fair degree of determinism can be imparted to such a search in
situations where the uniform provability property holds of a logical
language, and this fact has been utilized in the past in describing 
efficient proof procedures for suggested extensions to logic
programming; see, for instance, \cite{miller91jlc,Nad92int}. However,
there are logics of which the uniform provability property does not
hold directly. For example, suppose that our assumption set contains the
formula $p(a) \lor p(b)$ and that our desire is to prove
$\somex{x}p(x)$; our assumption set contains disjunctive information
in this case, typifying the situation in disjunctive logic
programming. We observe first that $\somex{x} p(x)$ is provable from
$p(a) \lor p(b)$ in classical, intuitionistic and minimal
logics. However, there is no uniform proof in any of these systems for
the given formula from the relevant assumption set; for such a proof
to exist, it is necessary that $p(t)$ be provable from the same
assumption set for some specific term $t$, and this requirement
clearly does not hold. As another
example, consider the formula $(p \supset q) \lor p$ in classical
logic. While this formula has a proof, it does not have a uniform one;
the latter kind of proof would exist only if either $(p \supset q)$ or $p$
is provable and, once again, clearly, neither is. The broad question
motivating the discussions in this paper is whether some benefit may
be derived from the uniform proof notion in structuring proof search
even in situations such as these where the uniform provability property
does not hold of the underlying logic.  

We answer this question below relative to classical logic. The main
observations we make are the following. Suppose that we wish to show
that a formula $G$ follows from a set of assumptions $\Gamma$ in
classical logic. We may not be able to do this immediately by looking
for a uniform proof. However, under a modest restriction in the syntax
of $\Gamma$ and $G$, there is a simple augmentation of
$\Gamma$ that makes the search for uniform proofs a complete
strategy. In particular, we show that if universal quantifiers do not
occur positively in $G$  or negatively in $\Gamma$, then there is a
proof for $G$ from $\Gamma$ in classical logic if and only if there is
a uniform proof for $G$ from $\Gamma, (G \supset \bot)$. This result
is actually a strengthening of the one in \cite{NL95lics} in that this
``modified'' uniform provability property is shown to hold for
an extension of disjunctive logic programming that permits hypothetical
goals. 
We further note that all uses of the added formula can be factored
into certain derived rules. 
These observations are then used to describe a simplified proof system
for classical logic.\footnote{The presentation of this proof system
  assumes a syntactic transformation of formulas. As we note later, 
  the only indispensable aspect of this transformation is the
  elimination of essentially positive occurrences of universal
  quantifiers.}
The resulting proof system provides the basis for a proof procedure
that generalizes the one usually employed in logic programming towards 
dealing with all of classical logic. An interesting aspect of this
proof procedure is that its rule for ``backchaining'' incorporates
within it the restart mechanism of nH-Prolog \cite{Lov87iclp,Lov91jar} for
dealing with disjunctive information in assumption sets and the
mechanism with the same name of QNR-Prolog \cite{Gab85} for handling
hypotheticals in goals.

\section{Logical preliminaries}\label{sec:basics}

We will work within the framework of a first-order logic in this
paper. The logical symbols that we assume as primitive are $\top$,
$\bot$, $\land$, $\lor$, $\supset$, $\some$, and $\all$. The first two
symbols in this collection denote the tautologous and the contradictory
propositions, respectively. The symbol $\neg$ is not primitive to our
language, but it can be easily defined using other symbols that are
primitive: $\neg A$ can be thought of as an abbreviation for $(A
\supset \bot)$.  

\begin{figure*}[top]

\begin{center}
  \mbox{\inrulean{\sequent{B,B,\Gamma}{\Delta}}
    {\sequent{B,\Gamma}{\Delta}} {\mthcontl}} \qquad\qquad\qquad\qquad
  \mbox{\inrulean{\sequent{\Gamma}{\Delta,B,B}} {\sequent{\Gamma}{\Delta,B}} {\mthcontr}}  \end{center}

\medskip

\begin{center}
    \mbox{\inrulean{\sequent{\Gamma}{\Delta,\bot}}
             {\sequent{\Gamma}{\Delta,D}}
             {\mthbotr}}
\end{center}

\medskip

\begin{center}
    \mbox{\inrulean{\sequent{B,D,B \land D, \Gamma}{\Delta}}
             {\sequent{B\land D,\Gamma}{\Delta}}
             {\mthandl}}
    \qquad\qquad\qquad\qquad
    \mbox{\inrulebn{\sequent{\Gamma}{\Delta,B}}
             {\sequent{\Gamma}{\Delta,D}}
             {\sequent{\Gamma}{\Delta,B\land D}}
             {\mthandr}}
\end{center}

\medskip

\begin{center}
   \mbox{
    \inrulebn{\sequent{B,\Gamma}{\Delta}}
             {\sequent{D,\Gamma}{\Delta}}
             {\sequent{B\lor D,\Gamma}{\Delta}}
             {\mthorl}
        }
\end{center}

\medskip

\begin{center}
    \mbox{\inrulean{\sequent{\Gamma}{\Delta,B}}
             {\sequent{\Gamma}{\Delta,B\lor D}}
             {\mthorr}}
    \qquad\qquad\qquad
    \mbox{\inrulean{\sequent{\Gamma}{\Delta,D}}
             {\sequent{\Gamma}{\Delta,B\lor D}}
             {\mthorr}}
\end{center}

\medskip

\begin{center}
    \mbox{\inrulebn{\sequent{B \supset D, \Gamma}{B, \Delta}}
             {\sequent{D,\Gamma}{\Theta}}
             {\sequent{B\supset D,\Gamma}{\Delta,\Theta}}
             {\mthimpl}}
    \qquad\qquad\qquad\qquad
    \mbox{\inrulean{\sequent{B,\Gamma}{\Delta,D}}
             {\sequent{\Gamma}{\Delta, B\supset D}}
             {\mthimpr}}
\end{center}

\medskip

\begin{center}
    \mbox{\inrulean{\sequent{\subs{t}{x} B,\allx{x} B,\Gamma}{\Delta}}
             {\sequent{\allx{x} B,\Gamma}{\Delta}}
             {\mthalll}}
    \qquad\qquad\qquad
    \mbox{\inrulean{\sequent{\Gamma}{\Delta, \subs{t}{x}B}}
             {\sequent{\Gamma}{\Delta, \somex{x} B}}
             {\mthsomer}}
\end{center}

\medskip

\begin{center}
    \mbox{\inrulean{\sequent{\subs{c}{x}B,\Gamma}{\Delta}}
             {\sequent{\somex{x}B,\Gamma}{\Delta}}
             {\mthsomel}}
    \qquad\qquad\qquad
    \mbox{\inrulean{\sequent{\Gamma}{\Delta,\subs{c}{x}B}}
             {\sequent{\Gamma}{\Delta,\allx{x} B}}
             {\mthallr}}
\end{center}

\medskip

\caption{Rules for Deriving Sequents \label{fig:seqrules}}
\end{figure*}

Notions of derivation that are of interest to us are formalized by
sequent calculi. A sequent in our context is a pair of multisets
of formulas. Assuming that $\Gamma$ and $\Delta$ are
its elements, the pair is written as $\sequent{\Gamma}{\Delta}$ and
$\Gamma$ and $\Delta$ are referred to as its antecedent and succedent,
respectively. Such a sequent is an axiom if either $\top \in \Delta$
or for some $A$ that is either $\bot$ or an atomic
formula,\footnote{The logical constants $\top$ and $\bot$ are not
  considered atomic formulas under our definition.} it is the
case that $A \in \Gamma$ and $A \in \Delta$. The rules that may be
used in constructing sequent proofs are those that can be obtained
from the schemata shown in Figure~\ref{fig:seqrules}. In these
schemata, $\Gamma$, $\Delta$ and $\Theta$ stand for multisets of
formulas, $B$ and $D$ stand for formulas, $c$ stands for a constant,
$x$ stands for a variable and $t$ stands for a term. The notation $B,
\Gamma$ ($\Delta, B$) is used here for a multiset containing the
formula $B$ whose remaining elements form the multiset $\Gamma$
(respectively, $\Delta$). Further, expressions of the form
$\subs{t}{x}B$ are used to denote the result of replacing all free
occurrences of $x$ in $B$ by $t$, with bound variables being renamed
as needed to ensure the logical correctness of these
replacements. There is the usual proviso with respect to the rules
produced from the schemata \somel\ and
\allr: the constant that replaces $c$ should not appear in the
formulas that form the lower sequent. The purpose of the schemata 
\contl\ and \contr\ is to blur the distinction between sets and
multisets, and so we will be ambivalent about this difference at 
times.

We are interested in three notions of derivability for sequents
of the form $\sequent{\Gamma}{B}$. A \Cbf-proof for such a sequent is
a derivation obtained by making arbitrary uses of the inference
rules. We denote the existence of such a proof, which is a
classical proof, for the sequent by writing $\Gamma \cprove B$. 
\Ibf-proofs, that formalize the notion of intuitionistic derivability,
are \Cbf-proofs in which every sequent has exactly one formula in its
succedent. We write $\Gamma \iprove B$ to indicate the existence of an
\Ibf-proof for $\sequent{\Gamma}{B}$. Finally, a 
{\it uniform proof}\ is an \Ibf-proof in which any sequent whose
succedent contains a 
non-atomic formula occurs only as the lower sequent of an inference
rule that introduces the top-level logical symbol of that
formula. Notice that if $\sequent{\Gamma}{B}$ has a uniform proof,
then the following must be true with respect to this proof:

\begin{enumerate}

\item If $B$ is $C \land D$, then the sequent must be
inferred by \andr\ from $\sequent{\Gamma}{C}$ and
$\sequent{\Gamma}{D}$. 

\item If $B$ is $C\lor D$ then the sequent must be inferred by \orr\
from either $\sequent{\Gamma}{C}$ or $\sequent{\Gamma}{D}$.

\item If $B$ is $\somex{x}P$ then the sequent must be inferred by \somer\
from $\sequent{\Gamma}{\subs{t}{x}P}$ for some term $t$.

\item If $B$ is $C\supset D$ then the sequent must be inferred by \impr\ 
from $\sequent{C,\Gamma}{D}$.

\item If $B$ is $\allx{x}P$ then, for some constant $c$ that
does not occur in the given sequent, it must be the case that the
sequent is inferred by \allr\  from $\sequent{\Gamma}{\subs{c}{x}P}$.

\end{enumerate}

\noindent These properties permit the search for a uniform proof to
proceed in a goal-directed fashion with the top-level structure of the
goal, \ie, the formula being proved, controlling the next step in the
search at each stage. 

We shall write $\Gamma \oprove B$ to denote the existence of a uniform
proof for $\sequent{\Gamma}{B}$; the subscript ${\scriptstyle O}$ is
used in the symbol for this derivability relation to indicate its
role in clarifying an operational notion of semantics in the
programming context. Letting $\cal D$ and $\cal G$ denote
collections of formulas and $\vdash$ denote a chosen proof
relation, an abstract logic programming language is defined in
\cite{MNPS91} as a triple $\langle {\cal D}, {\cal G}, \vdash \rangle$
such that, for all finite subsets $\cal P$ of $\cal D$ and all $G \in
{\cal G}$, ${\cal P} \vdash G$ if and only if ${\cal P} \oprove G$. 
In the programming interpretation of such a triple, elements of ${\cal
  D}$ function as program clauses and elements of ${\cal G}$ serve as
queries or goals and we therefore refer to each of these as such. 

The \alll\ rule usually included in sequent calculi has the form 

\begin{center}
    \mbox{\inrulewjan{\sequent{\subs{t}{x} B,\Gamma}{\Delta}}
             {\sequent{\allx{x} B,\Gamma}{\Delta}}}
\end{center}

\noindent In our presentation, we have combined this version of the
rule with the application of a \contl\ rule. It is easily seen that
the various provability relations of interest are the same under
either version of the \alll\ rule. An analogous remark applies to the
\andl\ rule. A comment of some interest is that our presentation of
the \alll\ rule actually renders the \contl\ rule redundant. However,
we do not use this fact in this paper. 

Our final observation concerns the so-called {\it Cut} rule that has
the following form:

\begin{center}
    \mbox{\inrulewjan{\sequent{\Gamma_1}{B,\Delta_1}\qquad\qquad
             \sequent{B,\Gamma_2}{\Delta_2}}
             {\sequent{\Gamma_1,\Gamma_2}{\Delta_1,\Delta_2}}
             }
\end{center}

\noindent It is well-known that this rule is admissible with respect
to classical and intuitionistic provability, \ie, the same set of
sequents have derivations with and without this rule. We use this
fact in the next section.

\section{Relating classical and intuitionistic
provability}\label{sec:corresp}

In considering the issue of uniform provability, it is usually necessary
to distinguish between the sets of logical symbols that are permitted
to appear positively and negatively in formulas. 
This distinction is, in fact, at the heart of the difference between
the goals and program clauses in an abstract logic programming
language.  
Our interest in this paper is in collections of formulas in
classical logic that turn out not to define an abstract logic
programming language. 
However, it is still useful to present the language that is of
interest to us using the vocabulary of goals and program clauses. 
This language is, in fact, the one in which these respective classes
of formulas are given by the syntax rules 
\[ \begin{array}{rcl}
 G & ::= & \top \sep \bot \sep A \sep G\land G \sep G\lor G \sep D
 \supset  G \sep \somex{x} G \\
 D & ::= & \top \sep \bot \sep A \sep G\supset D \sep D \land D \sep D
 \lor D \sep \somex{x} D \sep \allx{x} D \end{array} \]
in which $A$ represents an atomic formula. 
The collections described by these rules deviate from the set of all
formulas in that universal quantifiers are not permitted to appear
positively in $G$-formulas and negatively in $D$-formulas. 
However, there is a simple syntactic transformation that can be
applied to any given sequent to produce a new sequent whose antecedent
contains only $D$-formulas and whose succedent contains only
$G$-formulas and that is equivalent to the original sequent from the
perspective of classical provability; this transformation is the dual
of (static) Skolemization and is referred to as Herbrandization in
\cite{Shankar92}. 
The language presented above is also related at a syntactic level
to others that have been proposed previously. 
The logic of Horn clauses is obtained from it by not 
permitting
(a)~implications to appear as top-level symbols in $G$-formulas and
(b)~$\bot$, $\lor$ and $\some$ to appear as top-level symbols in
$D$-formulas. 
The language of hereditary  Harrop formulas \cite{MNPS91}
retains the second restriction but removes the first and, in addition,
permits universal quantifiers to appear as the top-level symbol in 
$G$-formulas. 
(The declarative  content of the resulting collections
of formulas is, in addition, clarified by intuitionistic
provability.) 
The N-clauses and N-goals of \cite{GaRe84} are subsumed by both the 
 $G$- and the $D$-formulas in the (restricted) language of hereditary
 Harrop formulas.
Finally, the logic underlying disjunctive logic programming
\cite{LMR92book,NL95lics} retains the $G$-formulas of Horn clause
logic but permits  $\lor$ and $\some$ to appear at the top-level in
$D$-formulas.  

We are ultimately interested in a uniform provability property for the
language described above. 
As a first step in this direction, we consider the relationship
between classical and intuitionistic provability for sequents of the
form $\sequent{\Gamma}{G}$ where $\Gamma$ is a collection of
$D$-formulas and $G$ is a $G$-formula. The category of
$G$-formulas includes a large subset of the formulas in first-order
logic, and so it is to be expected that these notions of provability
do not coincide for the sequents that are of interest. 
That this is in fact the case is seen by considering the sequent
$\sequent{}{((p \supset q) \supset p) \supset p}$; we assume here that
$p$ and $q$ are propositional symbols. As witnessed by the following
derivation, this sequent has a \Cbf-proof:
\begin{center}
\mbox{\inrulean
         {\inrulean
            {\inrulebn
                  {\linrulean
                        {\sequent{p, (p \supset q) \supset p}{q,p}}
                        {\sequent{(p \supset q) \supset p}{(p \supset q),
                            p}}
                        {\mthimpr}}
                  {\sequent{p}{p}}
                  {\sequent{(p \supset q) \supset p}{p,p}}
                  {\mthimpl}}
            {\sequent{(p \supset q) \supset p}{p}}
            {\mthcontr}}
         {\sequent{}{((p\supset q) \supset p) \supset p}}
         {\mthimpr}}
\end{center}
However, it is well-known that the sequent in question does not have
an \Ibf-proof. This situation is in contrast to the one that holds
in the context of most of the other mentioned languages whose
interpretation is based on classical logic: classical and intuitionistic 
provability are indistinguishable relative to the Horn
clause language \cite{MNPS91} and the language underlying
disjunctive logic programming \cite{NL95lics}.

The distinction between the two notions of provability
notwithstanding, there is a correspondence between the
classical provability of a sequent of the kind being considered and
the intuitionistic provability of a closely related sequent. In 
particular, a sequent of the form $\sequent{\Gamma}{G}$ has a
\Cbf-proof if and only if the sequent $\sequent{G \supset \bot,
  \Gamma}{G}$ has an \Ibf-proof. We establish this fact in this
section and use it later to extract a uniform provability property for
our language. 

We observe first that the mentioned augmentation of the set of
assumptions is one that is sound with respect to classical logic and,
in fact, without restrictions on the syntax of formulas.

\begin{lemma}\label{lem:equiv}
Let $\Gamma$ be a multiset of formulas and let $F$ be a formula. Then
$F \supset \bot, \Gamma \cprove F$ if and only if 
$\Gamma \cprove F$.  
\end{lemma}

\begin{proof}The if direction is obvious. For the only if direction,
  we note that $F, \Gamma \cprove F$ and so, if $F\supset \bot,
  \Gamma \cprove F$, then $F \lor (F \supset \bot), \Gamma \cprove
  F$. 
  Noting that $\cprove (F \supset \bot) \lor F$ and using the {\it
    Cut} rule, we see that $\Gamma \cprove F$. 
\end{proof}

Let $\Gamma$ represent a collection of $D$-formulas as defined above
and, similarly, let $G$ be a $G$-formula. 
In light of Lemma~\ref{lem:equiv}, the first step in the suggested
reduction of classical provability to uniform provability may
be justified by showing that a sequent of the form $\sequent{G 
  \supset \bot, \Gamma}{G}$ has a \Cbf-proof if and only if it has an
\Ibf-proof. It is this course that we follow below. Anticipating this
conclusion, we observe that Lemma~\ref{lem:equiv} cannot be true if the
relation $\cprove$ is replaced in it by $\iprove$ even in our
restricted context for otherwise the sequent
$\sequent{}{((p\supset q) \supset p) \supset p}$ would be
intuitionistically provable.

\begin{definition}\label{def:ncmeasure} Let $\Xi$ be a \Cbf-proof. 

\begin{enumerate}
\item An inference rule of the form 
\begin{center}
\mbox{\inrulewjan
         {\sequent{B,\Gamma}{\Delta} \qquad\qquad \sequent{D,\Gamma}{\Delta}}
         {\sequent{B \lor D,\Gamma}{\Delta}}}
\end{center}
\noindent that appears in $\Xi$ is said to be a nonconstructive
occurrence of an \orl\ rule just in case there is no $F$ in $\Delta$
such that $\sequent{B, \Gamma}{F}$ and $\sequent{D, \Gamma}{F}$ have
\Ibf-proofs. 

\item An inference rule of the form 
\begin{center}
\mbox{\inrulewjan
         {\sequent{B,\Gamma}{\Delta,D}}
         {\sequent{\Gamma}{\Delta,B \supset D}}}
\end{center}
\noindent that appears in $\Xi$  is said to be a nonconstructive
occurrence of an \impr\ rule just in case $\sequent{B,\Gamma}{D}$ does
not have an \Ibf-proof. 
\end{enumerate}

\noindent The nonconstructiveness measure of $\Xi$, denoted by
$\mu(\Xi)$, is the number of nonconstructive occurrences of \orl\ and
\impr\ rules in $\Xi$. 

\end{definition}

The following lemma explains the reason for singling out the \orl\ and
\impr\ rules and also casts light on the terminology of
Definition~\ref{def:ncmeasure}. 

\begin{lemma}\label{lem:sansorlimpr}If the sequent
  $\sequent{\Gamma}{\Delta}$ has a \Cbf-proof with
  nonconstructiveness measure $0$, then there is some formula $F \in
  \Delta$ such that $\sequent{\Gamma}{F}$ has an \Ibf-proof. 
\end{lemma}

\begin{proof}By an induction on the height of \Cbf-proofs.
\end{proof}

A converse to Lemma~\ref{lem:sansorlimpr} also holds. We state this
below in a more general form that is useful in subsequent
discussions. Note that an \Ibf-proof is a \Cbf-proof whose
nonconstructiveness measure is $0$.

\begin{lemma}\label{lem:conv}Let $\Gamma$ and $\Delta$ be multisets of
  formulas that are sub(multi)sets of $\Gamma'$ and $\Delta'$
  respectively. If $\sequent{\Gamma}{\Delta}$ has a \Cbf-proof of
  nonconstructiveness measure $n$, then $\sequent{\Gamma'}{\Delta'}$
  has a \Cbf-proof of nonconstructiveness measure $n$ or less.
\end{lemma}

\begin{proof}By an induction on the height of the \Cbf-proof of
  $\sequent{\Gamma}{\Delta}$. The
  essential idea is to show that the sequents in the \Cbf-proof of
  $\sequent{\Gamma}{\Delta}$ can be ``padded'' with new formulas while
  preserving the applicability of the inference
  rules. The constants used in some of the \somel\ and \allr\ rules
  may have to be ``renamed''
  to facilitate this, but it is easily seen that this can be done
  without altering the height or the nonconstructiveness measure of
  the derivation. 
\end{proof}

We show the main result of this section by arguing that there can
be no really nonconstructive occurrence of the rules \orl\ and \impr\
in a proof of a sequent of the form $\sequent{G \supset \bot,
  \Gamma}{G}$, where $G$ is a $G$-formula and $\Gamma$ is a multiset
of $D$-formulas. Towards this end, we develop machinery for transforming
apparently nonconstructive occurrences of the mentioned rules into
transparently ``constructive'' ones. 

\begin{definition}
We define an ordering on formulas that is intended to measure their
strength as assumptions: $F_1 \succeq F_2$ just in case $F_1 = F_2$ or
\begin{enumerate}
\item $F_2$ is $A \supset B$ and $F_1 \succeq B$,

\item $F_2$ is $A \lor B$ and $F_1 \succeq A$ or $F_1 \succeq B$, or

\item $F_2$ is $\somex{x}P$ and, for some constant $c$, $F_1 \succeq
  \subs{c}{x} P$. 
\end{enumerate}
\noindent This ordering is extended to multisets of formulas:
$\Gamma_1 \succeq \Gamma_2$ just in case there is a 1-1 mapping
$\kappa: \Gamma_2 \mapsto \Gamma_1$ such that $\kappa(F) \succeq F$. 
\end{definition}

\begin{lemma}\label{lem:getsstronger}
If $\sequent{\Gamma_1}{\Delta_1}$ and $\sequent{\Gamma_2}{\Delta_2}$
are two sequents appearing along a common path in a \Cbf-proof
(\Ibf-proof)  with the first appearing before the second, then
$\Gamma_1  \succeq \Gamma_2$. 
\end{lemma}

\begin{proof} By induction on the distance between the two sequents
  and an examination of the inference rules.
\end{proof}

\begin{lemma}\label{lem:fromstronger}
Let $\Gamma$ and $\Gamma'$ be two multisets of formulas such that
$\Gamma' \succeq \Gamma$. For any formula $F$, $\sequent{\Gamma}{F}$
has an \Ibf-proof only if $\sequent{\Gamma'}{F}$ has one. For any
multiset $\Delta$ of formulas, if $\sequent{\Gamma}{\Delta}$ has a
\Cbf-proof $\Xi$, then $\sequent{\Gamma'}{\Delta}$ has a \Cbf-proof
whose nonconstructiveness measure is at most that of
$\Xi$. 
\end{lemma}

\begin{proof}The essential idea is to construct a proof of
  $\sequent{\Gamma'}{F}$ ($\sequent{\Gamma'}{\Delta}$) by mimicking
  the given proof of   $\sequent{\Gamma'}{F}$
  ($\sequent{\Gamma'}{\Delta}$), possibly dropping some \impl, \orl\
    and \somel\ rules and thereby also pruning some branches. At a
    level of detail, we use an induction on the height of the given
    proof, showing the claim about \Ibf-proofs first and then using 
    this relative to ``constructive'' uses of \orl\ and \impl\ rules
    in proving the claim about \Cbf-proofs. 
\end{proof}

\begin{lemma}\label{lem:impr}
Let $\Gamma$ and $\Delta$ be multisets of $D$- and $G$-formulas
respectively. Further, let $\sequent{\Gamma}{\Delta}$ have a
\Cbf-proof $\Xi$ in which an \impr\ rule of the form
\begin{center}
\mbox{\inrulewjan
          {\sequent{B,\Sigma}{\Pi,D}}
          {\sequent{\Sigma}{\Pi,B \supset D}}}
\end{center}
\noindent occurs with the following characteristic: $\sequent{B,
  \Sigma}{D}$ does not have an \Ibf-proof but for some $F \in \Pi$,  
$\sequent{B, \Sigma}{F}$ has an \Ibf-proof. Then
$\sequent{B,\Sigma}{\Delta}$ has a \Cbf-proof whose nonconstructiveness
measure is smaller than that of $\Xi$. 
\end{lemma}

\begin{proof}By induction on the height of $\Xi$. At least one
  inference rule must have been used in $\Xi$. We consider first
  the possibility that the last such rule pertains to a formula in the
  antecedent and then that it pertains to a formula in the succedent.

The argument in the case of antecedent rules that have only one upper
sequent --- \ie, in the case of the rules \contl, \andl, \somel\ and
\alll --- takes a common form. In all these cases, the proof at the
end has the structure 
\begin{center}
\mbox{\inrulewjan
          {\sequent{\Gamma'}{\Delta}}
          {\sequent{\Gamma}{\Delta}}}
\end{center}
It is easily seen that all the formulas in $\Gamma'$ must be
$D$-formulas if those in $\Gamma$ are. Further, the \impr\ rule
mentioned in the lemma appears in the proof of
$\sequent{\Gamma'}{\Delta}$. Thus, the induction hypothesis can be
used to conclude that $\sequent{B,\Sigma}{\Delta}$ has a \Cbf-proof of
smaller nonconstructiveness measure than that of the proof of
$\sequent{\Gamma'}{\Delta}$. But the latter is actually identical to
$\mu(\Xi)$. 

Suppose that the last rule is an \orl, \ie, one of the form
\begin{center}
\mbox{\inrulewjan
          {\sequent{E,\Gamma'}{\Delta} \qquad\qquad
            \sequent{F,\Gamma'}{\Delta}}
          {\sequent{E \lor F, \Gamma'}{\Delta}}}
\end{center}
Now, the \impr\ rule mentioned in the lemma appears 
in the proof of either $\sequent{E,\Gamma'}{\Delta}$ or
$\sequent{F,\Gamma'}{\Delta}$. 
Without loss of generality, suppose the former. 
The induction hypothesis is again seen to be applicable
relative to the proof of $\sequent{E,\Gamma'}{\Delta}$. Using it and
noting that the nonconstructiveness measure of this proof is at most 
$\mu(\Xi)$ yields the desired conclusion.  

The only remaining possibility for an antecedent rule is \impl. In this
case, $\Xi$ has the form 
\begin{center}
\mbox{\inrulewjan
          {\sequent{E \supset F,\Gamma'}{\Delta_1, E} \qquad\qquad
            \sequent{F,\Gamma'}{\Delta_2}} 
          {\sequent{E \supset F, \Gamma'}{\Delta_1,\Delta_2}}}
\end{center}
at the end. The \impr\ rule mentioned in the lemma could appear
either above the left upper sequent or the right upper sequent of the
rule displayed. Suppose it is the latter. Noting that $F$ must be a
$D$-formula if $E\supset F$ is one and using the induction
hypothesis, we see that $\sequent{B,\Sigma}{\Delta_2}$ has a \Cbf-proof
of smaller nonconstructiveness measure than the one for
$\sequent{F,\Gamma'}{\Delta_2}$. The desired conclusion is now reached
by observing that the latter proof is a part of $\Xi$ and by employing
Lemma~\ref{lem:conv}. 

To complete the consideration of the case when an \impl\ is the last
rule used, suppose that the \impr\ rule mentioned in the lemma appears
above $\sequent{E \supset F,\Gamma'}{\Delta_1,E}$. Using the induction
hypothesis that is easily seen to be applicable, it follows that
$\sequent{B,\Sigma}{\Delta_1,E}$ has a \Cbf-proof whose 
nonconstructiveness measure is smaller than that of the given
\Cbf-proof of $\sequent{E \supset F,\Gamma'}{\Delta_1,E}$. By 
Lemma~\ref{lem:getsstronger}, $B, \Sigma \succeq (E \supset F),
\Gamma'$ and so $B, \Sigma$ can be written in the form $F', \Sigma'$
where $\Sigma' \succeq \Gamma'$ and either $F'$ is
identical to $E \supset F$ or $F' \succeq F$. In the former case, 
$F, \Sigma'  \succeq F, \Gamma'$ and so, by
Lemma~\ref{lem:fromstronger}, $\sequent{F, \Sigma'}{\Delta_2}$ has a
\Cbf-proof of nonconstructiveness measure at most that of the \Cbf-proof
of $\sequent{F, \Gamma'}{\Delta_2}$. Combining this with the
\Cbf-proof for $\sequent{B,\Sigma}{\Delta_1,E}$ yields one for
$\sequent{B,\Sigma}{\Delta_1,\Delta_2}$ with lower nonconstructiveness
measure than $\mu(\Xi)$. In the other case, \ie, when $F' \succeq F$,
it follows that $B, \Sigma \succeq F,\Gamma'$. Hence, by 
Lemmas~\ref{lem:fromstronger} and \ref{lem:conv}, 
$\sequent{F, B, \Sigma}{\Delta_2}$ has a \Cbf-proof with
nonconstructiveness measure at most that of
$\sequent{F,\Gamma'}{\Delta_2}$. By combining this \Cbf-proof
with that of $\sequent{B,\Sigma}{\Delta_1,E}$ we get one for
$\sequent{E \supset F, B, \Sigma}{\Delta_1,\Delta_2}$ that has a 
nonconstructiveness measure less than $\mu(\Xi)$. By 
Lemma~\ref{lem:fromstronger}, there is a \Cbf-proof with the same
characteristic for $\sequent{F', B,\Sigma}{\Delta_1,\Delta_2}$ and,
hence, using \contl, one for $\sequent{B, \Sigma}{\Delta_1,\Delta_2}$
as required. 

We now consider the possibilities for a succedent rule being the last one
in $\Xi$. The restriction in the syntax of the formulas in $\Delta$
ensures that this rule cannot be an \allr. If the last
rule is one of \botr, \orr, \somer\ and \contr, the same rule 
could be the last one in a purported \Cbf-proof of $\sequent{B,
  \Sigma}{\Delta}$ as well. Further the upper sequent of such a rule
application bears a 
relationship to the upper sequent of the corresponding rule
application in the \Cbf-proof of $\sequent{\Gamma}{\Delta}$ that
permits the induction hypothesis to be used. The desired conclusion
follows easily from these observations in these cases. 

An argument similar to the one for the succedent rules considered above
can also be provided in the case that the last rule in $\Xi$ is an
\andr. The only possibility that remains to be considered, then, is
that when an \impr\ rule is the last one. Here there are two subcases
to contend 
with: this rule may or may not be be the one mentioned in the
lemma. In the first situation, by Lemma~\ref{lem:conv},
$\sequent{B,\Sigma}{\Delta}$ has a \Cbf-proof whose nonconstructiveness
measure is $0$ and hence certainly less than $\mu(\Xi)$. In the other
situation, an argument similar to that for the other succedent rules
with a single upper sequent can be provided to show that
$\sequent{B,\Sigma}{\Delta}$ has a \Cbf-proof of nonconstructiveness
measure less than $\mu(\Xi)$.  

All the relevant cases having been considered, it follows that the
lemma must be true. 
\end{proof}

\begin{lemma}\label{lem:orl}
Let $\Gamma$ and $\Delta$ be multisets of $D$- and $G$-formulas
respectively. Further, let $\sequent{\Gamma}{\Delta}$ have a
\Cbf-proof $\Xi$ in which an 
\orl\ rule of the form
\begin{center}
\mbox{\inrulewjan
          {\sequent{B,\Sigma}{\Pi} \qquad\qquad \sequent{D,\Sigma}{\Pi}}
          {\sequent{B \lor D, \Sigma}{\Pi}}}
\end{center}
\noindent occurs with the following characteristic: there is no $F \in
\Pi$ such that $\sequent{B \lor D,\Sigma}{F}$ has an 
\Ibf-proof but there is an $F \in \Pi$ such that $\sequent{D,
  \Sigma}{F}$ has an \Ibf-proof. Then $\sequent{D,\Sigma}{\Delta}$ has
a \Cbf-proof whose nonconstructiveness measure is smaller than that of
$\Xi$. 
\end{lemma}

\begin{proof}By an argument similar to that for Lemma~\ref{lem:impr}.
\end{proof}

The restriction in the syntax of $D$- and $G$-formulas is
essential to the truth of Lemmas~\ref{lem:impr} and \ref{lem:orl}. For
instance, consider the following \Cbf-proof of
\[\sequent{}{\allx{x}q(x) \lor \somex{x}(q(x) \supset \bot)},\]
  assuming that $q$ represents a unary predicate symbol in this sequent: 

\sbox{\pfig}{\inrulean  
                    {\inrulean
                       {\inrulean
                           {\sequent{q(c)}{q(c),\bot}}
                           {\sequent{}{q(c), q(c) \supset \bot}}
                           {\mthbotr}}
                       {\sequent{}{q(c), \somex{x}(q(x)
                           \supset \bot)}}  
                       {\mthsomer}}
                    {\sequent{}{\allx{x}q(x), \somex{x}(q(x)
                             \supset \bot)}} 
                    {\mthallr}}

\begin{center}
\mbox{\inrulean
         {\inrulean
            {\inrulean
                {\usebox{\pfig}}
                {\sequent{}{\allx{x}q(x), \allx{x}q(x) \lor
                    \somex{x}(q(x) \supset \bot)}}
                {\mthorr}}
            {\sequent{}{\allx{x}q(x) \lor \somex{x}(q(x) \supset
               \bot), \allx{x}q(x) \lor \somex{x}(q(x) \supset \bot)}}
            {\mthorr}}
         {\sequent{}{\allx{x}q(x) \lor \somex{x}(q(x) \supset \bot)}}
         {\mthcontr}}
\end{center}

It is easily seen that $\sequent{q(c)}{\allx{x}q(x) \lor
    \somex{x}(q(x) \supset \bot)}$ does not have an \Ibf-proof as
  would be needed if Lemma~\ref{lem:impr} were to hold without
  restrictions. A similar observation can be made relative to
  Lemma~\ref{lem:orl} using the sequent $\sequent{\allx{x}(p \lor
    q(x))}{(p \lor \allx{x}q(x))}$ in which $p$ is assumed to be a
  proposition symbol and $q$ a unary predicate symbol. 

\begin{lemma}\label{lem:iproofexists}
Let $\Gamma$ be a multiset of $D$-formulas and let $G$ be a
$G$-formula such that \[\sequent{G \supset \bot, \Gamma}{G}\] has a
\Cbf-proof. If $\sequent{\Sigma}{\Pi}$ is a sequent that appears in
this proof, then there is some $F \in \Pi$ such that
$\sequent{\Sigma}{F}$ has an \Ibf-proof. 
\end{lemma}

\begin{proof}Suppose that the lemma is not true. Let $\Xi$ be a
  \Cbf-proof for a sequent of the form $\sequent{G\supset
    \bot,\Gamma}{G}$ that falsifies the lemma and, further, let $\Xi$
  have the smallest nonconstructiveness measure amongst
  \Cbf-proofs with this characteristic. 
  By Lemma~\ref{lem:sansorlimpr}, $\mu(\Xi)$ cannot be $0$. If
  $\mu(\Xi)$ is
  nonzero, there must be an \orl\ or an \impr\ rule in $\Xi$  that
  is the first nonconstructive occurrence of a rule of either kind
  along a branch. We consider each possibility below. 

Suppose that the rule in question is an \orl\ rule of the form 
\begin{center}
\mbox{\inrulewjan
         {\sequent{B,\Sigma}{\Pi} \qquad\qquad \sequent{D,\Sigma}{\Pi}}
         {\sequent{B \lor D,\Sigma}{\Pi}}}
\end{center}
By assumption, for no $F \in \Pi$ is it the case that an \Ibf-proof
exists for both $\sequent{B,\Sigma}{F}$ and $\sequent{D,\Sigma}{F}$. 
However, by Lemma~\ref{lem:sansorlimpr} and our assumption concerning
the structure of $\Xi$ prior to this rule, there must be some $F, F'
\in \Pi$ such that $\sequent{B,\Sigma}{F}$ and
$\sequent{D,\Sigma}{F'}$ have \Ibf-proofs. From the latter, using
Lemma~\ref{lem:orl}, it follows that $\sequent{D,\Sigma}{G}$ has a
\Cbf-proof with smaller nonconstructiveness measure than $\mu(\Xi)$.
Noting that the antecedent of every sequent in a \Cbf-proof of 
$\sequent{G\supset \bot,\Gamma}{G}$ must be a multiset of $D$-formulas
and then using the leastness assumption pertaining to $\mu(\Xi)$, we
conclude that $\sequent{D,\Sigma}{G}$ has an \Ibf-proof.
From Lemma~\ref{lem:getsstronger} and the fact that
$\sequent{B \lor   D,\Sigma}{\Pi}$  appears in a \Cbf-proof of
$\sequent{G\supset \bot,\Gamma}{G}$, it follows that $\Sigma$ is
either of the form $G\supset \bot,\Sigma'$ or of the form
$\bot,\Sigma'$. We assume the former, noting that the argument is
simpler if the latter is true.
  Now, we can construct the
following subderivation:
\begin{center}
\mbox{
        \inrulebn
            {\sequent{D,\Sigma}{G}}
            {\inrulean
                 {\sequent{\bot,\Sigma'}{\bot}}
                 {\sequent{\bot,\Sigma'}{F}}
                 {\mthbotr}}
            {\sequent{D,\Sigma}{F}}
            {\mthimpl}}
\end{center}
Using the \Ibf-proof that exists for $\sequent{D,\Sigma}{G}$ together
with this, we can obtain an \Ibf-proof for the sequent $\sequent{
  D,\Sigma}{F}$. But this is obviously a contradiction.  

Suppose instead that the rule of interest was a \impr\ rule of the
form 
\begin{center}
\mbox{\inrulewjan
          {\sequent{B,\Sigma}{\Pi,D}}
          {\sequent{\Sigma}{\Pi,B \supset D}}}
\end{center}
\noindent By our assumptions and Lemma~\ref{lem:sansorlimpr}, we have the
following: $\sequent{B,\Sigma}{D}$ does not have an \Ibf-proof, but
for some $F \in \Delta$ it is the case that $\sequent{B,\Sigma}{F}$
has an \Ibf-proof. From the latter and Lemma~\ref{lem:impr} it follows
that $\sequent{B,\Sigma}{G}$ has a \Cbf-proof whose 
nonconstructiveness measure is less than $\mu(\Xi)$. We can, once
again, conclude from this that $\sequent{B,\Sigma}{G}$ has
an \Ibf-proof. From Lemma~\ref{lem:getsstronger} it follows that
$\Sigma$ can be written as either $G \supset \bot,\Sigma'$ or
$\bot,\Sigma'$. We assume the former, noting as before that the
argument becomes simpler if the latter is true. Now, the following
subderivation can be constructed:  
\begin{center}
\mbox{
           \inrulebn
              {\sequent{B,\Sigma}{G}}
              {\inrulean
                   {\sequent{\bot,B,\Sigma'}{\bot}}
                   {\sequent{\bot,B,\Sigma'}{D}}
                   {\mthbotr}}
              {\sequent{B,\Sigma}{D}}
              {\mthimpl}}
\end{center}
\noindent Using the \Ibf-proof of $\sequent{B,\Sigma}{G}$ together
with this, we obtain an \Ibf-proof for the sequent
$\sequent{B,\Sigma}{D}$, yielding, once again, a contradiction. 

It is thus untenable that the lemma is false and so it must, in fact,
be true.  
\end{proof}

The main conclusion that we desire in this section is an easy
corollary of Lemmas~\ref{lem:equiv} and \ref{lem:iproofexists}.

\begin{theorem}\label{thm:equiv}
Let $\Gamma$ be any collection of $D$-formulas and let $G$ be a
$G$-formula. Then $\Gamma \cprove G$ if and only if $G \supset \bot,
\Gamma \iprove G$.  
\end{theorem}

The restriction in the syntax of $D$- and $G$-formulas is important to
the truth of Theorem~\ref{thm:equiv}, a fact that we became aware of
through the comments of Robert St\"{a}rk. To see that this is the case,
consider the sequent $\sequent{\allx{x}((p(x) \supset \bot) \supset
  \bot)}{\allx{x}p(x)}$ in which $p$ is a unary predicate symbol. This
sequent has a \Cbf-proof but the sequent \[\sequent{(\allx{x}p(x)
  \supset \bot), \allx{x}((p(x) \supset \bot) \supset
  \bot)}{\allx{x}p(x)}\]  does {\it not} have an \Ibf-proof. A
question of interest is whether the theorem can be strengthened in
any way. In particular, are there alternative restrictions that can be
placed on the syntax of $G$ and the formulas in $\Gamma$ that do not
presuppose specific knowledge of these formulas but still ensure that
$\sequent{(G \supset \bot),\Gamma}{G}$ has an \Ibf-proof whenever
$\sequent{\Gamma}{G}$ has a \Cbf-proof? In response to this question,
we note that this assurance can be given under only two circumstances:
when the syntactic restrictions guarantee that $\sequent{\Gamma}{G}$
itself has an \Ibf-proof (and these restrictions do not always
preclude the use of the \allr\ rule) and when they ensure that the
\allr\ rule will not be utilized. Thus, the restrictions assumed in
Theorem~\ref{thm:equiv} reflect the most liberal ones that allow
classical provability to be reduced to intuitionistic provability
through the indicated augmentation to the assumption set and where
this reduction is a non-trivial one. A detailed discussion of these
and other matters is planned for a sequel to this paper. 

The proofs of the various lemmas in this section, culminating in that
of Lemma~\ref{lem:iproofexists}, contain more information 
than is utilized in proving Theorem~\ref{thm:equiv}. One particular
aspect that we note here is their constructive content: under a
suitable interpretation, they provide the basis for 
a procedure that takes a \Cbf-proof for a sequent of the 
form $\sequent{\Gamma}{G}$ and, by working downward from the leaves in
this proof, that transforms this into an \Ibf-proof for $\sequent{G
  \supset  \bot, \Gamma}{G}$. 
For example, consider the \Cbf-proof for $\sequent{}{((p \supset q)
  \supset p) \supset p}$ displayed at the beginning of this section.
Assuming that $G \supset \bot$ denotes the formula 
\[(((p \supset q) \supset p) \supset p) \supset \bot,\] 
the mentioned procedure would transform this \Cbf-proof into the
following \Ibf-proof for a suitably augmented sequent:  

\sbox{\partfig}
             {\inrulebn
                     {\linrulean
                        {\sequent{(p  \supset q) \supset p, p, (p
                            \supset q) \supset p, 
                            G \supset \bot}{p}}
                        {\sequent{p, (p  \supset q) \supset p, 
                            G \supset \bot}{((p  \supset q) \supset
                            p) \supset 
                            p}}
                        {\mthimpr}}
                     {\rinrulean
                        {\sequent{\bot,p, (p  \supset q) \supset p}{\bot}}
                        {\sequent{\bot, p, (p  \supset q) \supset p}{q}}
                        {\mthbotr}}
                     {\sequent{p, (p  \supset q) \supset p, 
                        G \supset \bot}{q}}
                     {\mthimpl}}

\begin{center}
\mbox{\hspace{3.0cm}
      \inrulean
        {\inrulebn
             {\linrulean
                 {\usebox{\partfig}}
                 {\sequent{(p  \supset q) \supset p, 
                           G \supset \bot}{p \supset q}}
                 {\mthimpr}}
             {\sequent{p,
                 G \supset \bot}{p}} 
             {\sequent{(p  \supset q) \supset p, 
                 G \supset \bot}{p}}
             {\mthimpl}}
        {\sequent{
                 G \supset \bot}{((p  \supset q) \supset p) \supset p}}
        {\mthimpr}}
\end{center}

\noindent This observation can be further sharpened by noting that
only very restricted uses are made of the added formula in the
transformation process. We utilize this fact in the next section in
describing a modified deductive calculus for our language in which the
augmentation of sequents is made implicit.

\section{A uniform provability property}\label{sec:upprop}

The uniform provability property fails to hold in an immediate sense
for our fragment of classical logic. 
This is not a surprising fact, given that intuitionistic provability
is already a more restrictive relation than classical provability
relative to our language. Furthermore, intuitionistic provability is
itself distinct in this context from uniform provability. This latter
difference arises from the possibility for disjunctive and existential
information to be present in assumptions. Thus, consider the sequent
$\sequent{p(a) \lor   p(b)}{\somex{x}p(x)}$. This sequent has the
following \Ibf-proof: 
\begin{center}
\mbox{\inrulebn{\linrulean{\sequent{p(a)}{p(a)}}
                          {\sequent{p(a)}{\somex{x}p(x)}}
                          {\mthsomer}}
               {\rinrulean{\sequent{p(b)}{p(b)}}
                          {\sequent{p(b)}{\somex{x}p(x)}}
                          {\mthsomer}}
               {\sequent{p(a) \lor p(b)}{\somex{x}p(x)}}
               {\mthorl}
\hspace{0.5cm}}
\end{center}
However, as already noted, there can be no uniform proof for this sequent. 

While the uniform provability property does not hold in a strict sense
for our fragment of classical logic, it does hold of it in a
derivative sense: assuming that $\Gamma$ is a set of
$D$-formulas and $G$ is a $G$-formula, a \Cbf-proof exists for
$\sequent{\Gamma}{G}$ if and only if a uniform proof exists for
$\sequent{G\supset \bot, \Gamma}{G}$. In the previous 
section, we have already observed that the indicated augmentation of
the assumption set yields a correspondence between classical and
intuitionistic provability. Thus, one way to establish the above uniform
provability property is to show that the same augmentation also leads
to a coincidence between intuitionistic and uniform provability. 
A proof of this fact relative to the logic underlying disjunctive
logic programming is provided in \cite{NL95lics} and it turns out that
this argument can be extended to the present context as
well. We do this below, taking care to cast our discussions in a
form that supports a subsequent extraction from them of a proof
procedure for classical logic. 

Our first step in the indicated direction is to refine the deductive
calculus to be used for constructing derivations for the kinds of
sequents of interest to us. 
In particular, consider the following inference rules that are
parameterized by a specific formula $G$: 

\medskip

\begin{center}
\mbox{\inrulebn
         {\sequent{B,\Delta}{F}}
         {\sequent{D,\Delta}{G}}
         {\sequent{B \lor D, \Delta}{F}}
         {\mthorlg}}

\bigskip

\mbox{\inrulean
         {\sequent{\Delta}{G}}
         {\sequent{\Delta}{F}}
         {\mthresg}}
\end{center}

\medskip

\noindent We assume that $B$, $D$ and $F$ are schema variables for
formulas in these rules and that $\Delta$ denotes a multiset of
formulas. It is 
easily seen that these rules are derived ones relative to the
sequent calculus for intuitionistic logic in the case that $\Delta$
contains the formula $G \supset \bot$. 
Now, as noted at the end of the last section, the transformation
procedure implicit in the proof of Lemma~\ref{lem:iproofexists} yields
an \Ibf-proof for $\sequent{G \supset \bot, \Gamma}{G}$ in which {\it
  every} use that is made of the formula $G \supset \bot$ that is
added to the assumptions can be seen to be embedded within one of
these derived rules.  
Thus, by using these rules and by strengthening the proviso on the
\somel\ and \allr\ rules to disallow the use of constants appearing in
$G$, the augmentation of sequents can be made implicit. 

Let us tentatively refer to derivations constructed in a sequent calculus
obtained from that for intuitionistic logic through the above 
modifications as \Ibfg-proofs. We now make the following further
observation: the \orl\ rule is redundant from the perspective of
constructing \Ibfg-proofs for the kinds of sequents of
interest to us. This observation is a consequence of the lemma
below whose proof, when viewed constructively, provides the basis for
transforming \orl\ rule occurrences into occurrences of the \orlg\ rule. 

\begin{lemma}\label{lem:orltoorlg} Let $\Gamma$ be a multiset of
  $D$-formulas and let $G$ be a $G$-formula. Further, let
  $\sequent{\Gamma}{G}$ have an \Ibfg-proof $\Xi$ in which an \orl\
  rule of the form

\begin{center}
\mbox{\inrulewjan
          {\sequent{B,\Sigma}{F} \qquad\qquad \sequent{D,\Sigma}{F}}
          {\sequent{B \lor D, \Sigma}{F}}}
\end{center}

\noindent appears. Then $\sequent{D, \Sigma}{G}$ has an \Ibfg-proof
in which there are fewer occurrences of the \orl\ rule than in $\Xi$. 
\end{lemma}

\begin{proof} By induction on the height of the given \Ibfg-proof,
  using the analogue of Lemma~\ref{lem:getsstronger} for \Ibfg-proofs
  and the easily established fact   that if $\Delta' \succeq \Delta$
  and $\sequent{\Delta}{F}$ has an \Ibfg-proof with $n$ occurrences of
  the \orl\ rule, then $\sequent{\Delta'}{F}$ has an \Ibfg-proof with
  $n$ or fewer occurrences of the \orl\ rule. We omit the details of
  the argument, noting that they are similar to those in the proofs of
  Lemmas~\ref{lem:impr} and \ref{lem:orl}. 
\end{proof}

The restriction in the syntax of $D$- and $G$-formulas
is essential to the truth of the above lemma. Thus, consider
the following \Ibfg-proof for 
\[\sequent{}{\allx{x}((p(x,a) \lor p(x,b)) \supset
  \somex{y}p(x,y))},\]
assuming that $p$ is a binary 
predicate symbol and $a$, $b$ and $c$ are constant symbols:

\begin{center}
\mbox{\inrulean{\inrulean
                   {\inrulebn
                       {\inrulean
                           {\sequent{p(c,a)}{p(c,a)}}
                           {\sequent{p(c,a)}{\somex{y}p(c,y)}}
                           {\mthsomer}}
                       {\inrulean
                           {\sequent{p(c,b)}{p(c,b)}}
                           {\sequent{p(c,b)}{\somex{y}p(c,y)}}
                           {\mthsomer}}
                       {\sequent{(p(c,a) \lor
                           p(c,b))}{\somex{y}p(c,y)}}
                       {\mthorl}}
                   {\sequent{}{(p(c,a) \lor p(c,b))
                               \supset \somex{y}p(c,y)}}
                   {\mthimpr}}
               {\sequent{}{\allx{x}((p(x,a) \lor p(x,b))
                               \supset \somex{y}p(x,y))}}
               {\mthallr}
\hspace{0.5cm}}
\end{center}

\noindent This derivation has one occurrence of an \orl\ rule and it
is easily seen that there is no \Ibfg-proof for the sequent 
$\sequent{p(c,b)}{\allx{x}((p(x,a) \lor p(x,b)) \supset
  \somex{y}p(x,y))}$ in which there are no occurrences of the \orl\ rule.

On the strength of Lemma~\ref{lem:orltoorlg} and the comments
preceding it, we assume henceforth that the \Ibfg-proofs that we
consider {\it do not} contain occurrences of the \orl\ rule. The
results of the previous section can now be summarized in the context
of our present discussion as follows:

\begin{theorem}\label{thm:firstmodified}
Let $\Gamma$ be a multiset of $D$-formulas and let $G$ be a
$G$-formula. Then the sequent $\sequent{\Gamma}{G}$ has a \Cbf-proof
if and only if it has a \Ibfg-proof.
\end{theorem}

We now relativize the notion of a uniform proof to our modified
calculus. 
In particular, let an \Obfg-proof be an \Ibfg-proof with the following
characteristic: if there is a sequent in this proof whose succedent
contains a non-atomic formula, then that sequent occurs as the
lower sequent of an inference rule that introduces the top-level
logical symbol of that formula. 
The following may then be observed:

\begin{lemma}\label{lem:secondmodified}
Let $G$ be a $G$-formula and let $\Gamma$ be a multiset of
$D$-formulas. Then $\sequent{\Gamma}{G}$ has a \Ibfg-proof only if it
has an \Obfg-proof.  
\end{lemma}

\begin{proof}
Suppose that $\sequent{\Gamma}{G}$ has an
\Ibfg-proof. 
It must then have an \Ibfg-proof in which there is
a rule introducing the top-level logical symbol of every non-atomic
formula appearing in the succedent of a sequent; to ensure that
this is the case, we only need to introduce some inference
steps right after \botr\ and \resg\ rules in a manner whose details
are entirely transparent. Further, an \Ibfg-proof
of this kind exists that also satisfies the following additional
condition: no antecedent rule immediately succeeds a succedent rule
pertaining 
to a top-level logical symbol of a formula in a common sequent except 
in the case that the antecedent rule is \somel\ and the succedent rule
is \somer.  
To see that this is so, we first observe, by an easy induction on
the heights of \Ibfg-proofs, that (a)~if a sequent of the form
$\sequent{\Sigma}{A\land B}$ has an \Ibfg-proof of height $h$, then
both $\sequent{\Sigma}{A}$ and $\sequent{\Sigma}{B}$ have \Ibfg-proofs
of height $h$ or less, (b)~if a sequent of the form
$\sequent{\Sigma}{A \supset B}$ has an \Ibfg-proof of height $h$, then
$\sequent{A,\Sigma}{B}$ has an \Ibfg-proof of height $h$ or
less, and (c)~if a sequent of the form $\sequent{\Sigma}{\allx{x}B}$
has an \Ibfg-proof of height $h$, then, for any constant $c$,
$\sequent{\Sigma}{\subs{c}{x}B}$ has an \Ibfg-proof of height $h$ or
less.  
Now, given an \Ibfg-proof for a sequent of the form $\sequent{\Sigma}{F}$,
let us associate with this proof the pair of natural numbers $\langle
n_1, n_2 \rangle$ in which $n_1$ is the height of the given \Ibfg-proof
and $n_2$ is the count of the number of logical symbols in $F$ and let
us consider an ordering on \Ibfg-proofs that is based on the
the extension of the usual ordering on the natural numbers to a
lexicographic ordering on the pairs of numbers corresponding to the
proofs. The existence of an \Ibfg-proof of the required form is
established by an induction on the mentioned ordering. 

Let us call an \Ibfg-proof satisfying the requirements mentioned above
an \Ibfgprime-proof.  
We then define the {\it nonuniformity measure} of a \somel\ rule as
the count of the number of connectives and quantifiers that appear in
the succedent of the lower sequent of the rule, and the nonuniformity
measure of an \Ibfgprime-proof as the sum of the (nonuniformity)
measures of the  \somel\ rules that appear in it. We claim that any
\Ibfgprime-proof of $\sequent{\Gamma}{G}$ that has a
nonzero nonuniformity measure can be 
transformed into an \Ibfgprime-proof of smaller measure. It follows
from this that $\sequent{\Gamma}{G}$ has an \Obfg-proof. 

To show the claim, suppose that the \Ibfgprime-proof of
$\sequent{\Gamma}{G}$ in fact has a nonzero nonuniformity 
measure. It must then be the case that somewhere in the derivation an
\somel\ rule appears right after an \somer\ rule. In other words,
there is a subderivation of the form 
\begin{center}
\mbox{\inrulean{\inrulean
                  {\sequent{\subs{c}{x}B, \Gamma'}{\subs{t}{y}D}}
                  {\sequent{\subs{c}{x}B, \Gamma'}{\somex{y}D}}
                  {\mthsomer}}
               {\sequent{\somex{x}B, \Gamma'}{\somex{y}D}}
               {\mthsomel}}
\end{center}
at some point in the given \Ibfgprime-proof. 
Let us assume that $D$ is atomic --- this assumption is not really
essential and can be dispensed with in a more detailed 
argument. 
Using the fact that what is displayed above is a subpart of an
\Ibfgprime-proof of $\sequent{\Gamma}{G}$, it can be shown that
$\sequent{\subs{c}{x}B,\Gamma'}{G}$ has an \Ibfgprime-proof of smaller
nonuniformity measure than that of the one for $\sequent{\Gamma}{G}$; as in
the case of Lemma~\ref{lem:impr}, the essential idea is to mimic the
structure of the given proof of $\sequent{\Gamma}{G}$ while noting
that at least one occurrence of an \somel\ rule --- the one shown above
--- that makes a nonzero contribution to the nonuniformity measure
can be eliminated. By induction it follows then that
$\sequent{\subs{c}{x}B,\Gamma'}{G}$ has an \Obfg-proof. We further
observe that the proviso on a \somel\ rule ensures that $c$ does not
occur in $B$, $\Gamma'$ or $G$. From this it is easily seen, for any
constant $c'$, $\sequent{\subs{c'}{x}B,\Gamma'}{G}$ has an
\Obfg-proof. Let $c'$ be a constant that does not occur in $t$ in
addition to not appearing in $B$, $\Gamma'$ and $G$. Then we
can construct the following subderivation:
\begin{center}
\mbox{\inrulean{\inrulean
                  {\inrulean
                         {\sequent{\subs{c'}{x}B, \Gamma'}{G}}
                         {\sequent{\subs{c'}{x}B,
                             \Gamma'}{\subs{t}{y}D}}
                         {\mthresg}}
                  {\sequent{\somex{x}B, \Gamma'}{\subs{t}{y}D}}
                  {\mthsomel}}
               {\sequent{\somex{x}B, \Gamma'}{\somex{y}D}}
               {\mthsomer}}
\end{center}
Using the known \Obfg-proof for $\sequent{\subs{c'}{x}B,\Gamma'}{G}$
together with this to replace the earlier subderivation, we obtain the
desired \Ibfgprime-proof of reduced measure. 
\end{proof}

The syntactic restrictions on $D$- and $G$-formulas
are, once again, necessary for the truth of
Lemma~\ref{lem:secondmodified}: assuming that $p$ and $q$ are binary
predicate symbols, it can be seen, for instance, that the sequent
\[\sequent{\allx{x}\allx{y}p(x,y)}{\allx{x}((\somex{y}(p(x,y) \supset
  q(x,y))) \supset \somex{y}q(x,y))}\]
has an \Ibfg-proof but {\it does not} have an \Obfg-proof. 

The uniform provability property is an easy consequence of
Lemma~\ref{lem:secondmodified}. 

\begin{theorem}\label{thm:upprop}
Let $G$ be a $G$-formula and let $\Gamma$ be a multiset of
$D$-formulas. Then $\Gamma \cprove G$ if and only if $G \supset \bot,
\Gamma \oprove G$.  
\end{theorem}

\begin{proof}
Given Lemma~\ref{lem:equiv}, the if direction is obvious. For the only
if direction, we use Theorem~\ref{thm:firstmodified},
Lemma~\ref{lem:secondmodified} and the fact that an \Obfg-proof for
$\sequent{\Gamma}{G}$ can be translated into a uniform proof for
$\sequent{G \supset \bot, \Gamma}{G}$. 
\end{proof}

There is a constructive content to the proofs of
Lemmas~\ref{lem:orltoorlg} and ~\ref{lem:secondmodified} and it is
useful to understand this. For this purpose, consider the proof for
the sequent 
\[\sequent{p(a) \lor p(b)}{\somex{x}p(x)}\] that is shown at the
beginning of this section. Construing the inference rule labelled as
\orl\ as an \orlg\ rule instead, this proof is seen also to be an
\Ibfg-proof. Based on the argument provided for 
Lemma~\ref{lem:secondmodified}, this proof can be transformed into
the \Obfg-proof
\begin{center}
\mbox{\inrulean
         {\inrulebn{\sequent{p(a)}{p(a)}}
                   {\rinrulean{\sequent{p(b)}{p(b)}}
                          {\sequent{p(b)}{\somex{x}p(x)}}
                          {\mthsomer}}
                   {\sequent{p(a) \lor p(b)}{p(a)}}
                   {\mthorlg}}
         {\sequent{p(a) \lor p(b)}{\somex{x}p(x)}}
         {\mthsomer}
\hspace{0.5cm}}
\end{center}

\section{A reduced proof system for classical logic}\label{sec:proofproc}

The uniform provability property that was established 
in the previous section is useful in describing a proof
procedure for classical logic. 
The starting point for such a procedure is a formula from which
essentially positive occurrences of universal quantifiers have been
eliminated by the process of Herbrandization. 
Now, whenever the procedure is required to find a proof for a
non-atomic formula, it uses the top-level logical symbol in this
formula to determine the next step in proof search. 
However, the way to proceed is not quite so clear when an atomic
formula has been produced through this process: the
\botr\ rule and a variety of antecedent rules may be applicable at this
point and there is at present no mechanism for picking between
these. We outline an approach to dealing with this situation in this
section. 
This approach is based on combining the antecedent rules and the
\botr\ rule into a generalization of the backchaining rule that is
known from Horn clause logic and that, in a sense, is controlled by
the atomic formula for which a proof is sought. 
In our context there will be three different manifestations of this
rule, and, as is typically the case, more than one instance of these
forms of the rule might be applicable at a relevant stage in the proof
search process. 
The manner in which a choice is made between these different
possibilities  could have a substantial impact on the behavior of an
actual proof procedure. 
However, we stop short of considering the pragmatically important
question of how this choice is to be made, presenting only the basic
structure of the proof procedure through a reduced proof system.  

Our main objective, then, is the enunciation of a suitable
backchaining rule. 
In stating this rule and in manifesting its intuitive content, it is
preferable to use a simplified syntax for $D$- and $G$-formulas. 
In particular, we assume from now on that our goals and program
clauses are given by the rules 
\[ \begin{array}{rcl}
 G & ::= & A \sep G\land G \sep G\lor G \sep D
 \supset  G \sep \somex{x} G \\
 D & ::= & (A \lor \ldots \lor  A) \sep G\supset (A \lor \ldots \lor A)\sep
 \allx{x} D \end{array} \] 
in which the symbol $A$ is assumed to represent the category of atomic
formulas augmented by the logical constants $\top$ and $\bot$. 
Using known logical equivalences and the notion of (static)
Herbrandization, the question of classical provability of a 
sequent of the form $\sequent{\Gamma}{F}$ in which the formulas are
permitted to have an arbitrary syntax can be transformed into an
identical question for a similar sequent in which the assumption and
goal formulas adhere to the respective simplified syntax. The use of
this ``reduced'' syntax therefore does not 
constitute a loss of generality in our discussions.\footnote{With the
  exception of the elimination of certain occurrences of
  universal quantifiers, the simplification in the syntax of formulas
  is also not essential and is chosen mainly for reasons of
  perspicuity. 
  An alternative approach would be to incorporate the mentioned
  syntactic transformation of formulas implicitly into the definition
  of the instances of a program clause that follows. Notice, however,
  that the proper treatment of existential quantification in program
  clauses under this approach would require the relativization of the
  definition of clause instances to a given signature.}

Our backchaining rule will be based, as usual, on the
notion of an instance of a program clause. In
the present setting this notion is explicated as follows:

\begin{definition}\label{def:instances}
Let $D$ be a program clause. Then $[ D]$ denotes a collection of pairs
of sets of formulas given as follows:

\begin{enumerate}

\item If $D$ is $A_1 \lor \ldots \lor A_n$, then 
$[D] = \{\langle \emptyset, \{A_1,\ldots A_n\}\rangle\}$.

\item If $D$ is $G \supset (A_1\lor \ldots \lor A_n)$, then 
$[D] = \{\langle \{G\}, \{A_1,\ldots,A_n\}\rangle\}$.

\item If $D$ is $\allx{x}D_1$, then 
$[D] = \bigcup \{[\subs{t}{x}D_1] \sep t\ \hbox{\rm is a term}\}$.

\end{enumerate}

\noindent This notation is extended to a (multi)set $\Gamma$ of program
clauses as follows: \[[\Gamma] = \bigcup \{[D] \sep D \in \Gamma\}.\]
\end{definition}

The starting point for our proof search is represented by a sequent of
the form $\sequent{\Gamma}{G}$ in which $\Gamma$ is a multiset of
$D$-formulas and $G$ is a $G$-formula. In the discussions that
follow, we assume a calculus for \Obfg-proofs that is relativized to
this starting sequent; in particular, the $G$-formula in the \orlg\ and
\resg\ rules is chosen to coincide with its succedent formula.

The following lemma underlies our generalization of the backchaining
rule. We adopt a harmless abuse of notation in the statement of this
lemma and in the subsequent discussions in that we permit $n$ to be
$0$ in a listing $A_1,\ldots,A_n$  of formulas, assuming, in this
case, that the ``listing'' denotes an empty sequence. 

\begin{lemma}\label{lem:backchain}
Let $\Gamma$ be a multiset of program clauses and let $C$ be an atomic
formula or $\bot$. Then $\sequent{\Gamma}{C}$ has an \Obfg-proof with 
$l$ sequents appearing in it just in case one of the following holds:

\begin{enumerate}

\item $\sequent{\Gamma}{G}$ has an \Obfg-proof with fewer than $l$
  sequents in it.

\item For some $A_1,\ldots,A_n$, it is the case that either $\langle
  \emptyset, \{ C, A_1, 
  \ldots, A_n \} \rangle \in [\Gamma]$ or $\langle \emptyset, \{ \bot, A_1,
  \ldots, A_n \} \rangle \in [\Gamma]$ and, if $n > 0$ then, for $1
  \leq i \leq n$, $\sequent{A_i,\Gamma}{G}$ has an \Obfg-proof with
  fewer than $l$ sequents.

\item For some $G'$ and $A_1, \ldots, A_n$ it is the case that either
  $\langle \{G'\}, \{ C, A_1, 
  \ldots, A_n \} \rangle$ or  $\langle \{G'\}, \{ \bot, A_1,   \ldots,
  A_n \} \rangle$ is a member of $[\Gamma]$
  and $\sequent{\Gamma}{G'}$ and, if $n > 0$ then, for $1   \leq i \leq n$,
  $\sequent{A_i,\Gamma}{G}$ have \Obfg-proofs with 
  fewer than $l$ sequents.

\end{enumerate}

\end{lemma}

\begin{proof} An easy induction on the size of an \Obfg-proof for a
  sequent of the kind that is of interest.
\end{proof}

The content of the above lemma from the perspective of proof search is
abstracted into the following definition.

\begin{definition}\label{def:backchain}
Let $\Gamma$ represent a multiset of program clauses and let $C$
represent an atomic formula or $\bot$. We describe three rules
below that are relativized to a particular choice of goal 
$G$. 

\begin{enumerate}

\item The RESTART rule is the following:

\begin{center}
\mbox{\inrulewjan{\sequent{\Gamma}{G}}
                 {\sequent{\Gamma}{C}}
}
\end{center}

\item The ATOMIC rule is the following

\begin{center}
\mbox{\inrulewjan{\sequent{A_1,\Gamma}{G}\qquad\ldots\qquad\sequent{A_n,\Gamma}{G}}
                 {\sequent{\Gamma}{C}}
}
\end{center}

provided that $\langle \emptyset, \{C,A_1,\ldots,A_n\}\rangle \in
[\Gamma]$ or $\langle \emptyset, \{\bot,A_1,\ldots,A_n\}\rangle \in
[\Gamma]$. In the degenerate case, \ie, when the second component of
the pair shown is simply $\{C\}$ or $\{\bot\}$, this
rule has no upper sequents and, in this case, constitutes an axiom.

\item The BACKCHAIN rule is the following

\begin{center}
\mbox{\inrulewjan{\sequent{\Gamma}{G'}\quad\sequent{A_1,\Gamma}{G}
\quad \ldots\quad \sequent{A_n,\Gamma}{G}}
                 {\sequent{\Gamma}{C}}
}
\end{center}

provided that $\langle \{G'\} ,\{C,A_1,\ldots,A_n\}\rangle \in
[\Gamma]$ or $\langle \{G'\} ,\{\bot,A_1,\ldots,A_n\}\rangle \in
[\Gamma]$. In the degenerate case, \ie, when the second component of
the pair shown is simply $\{C\}$ or $\{\bot\}$, this rule has 
$\sequent{\Gamma}{G'}$ as its only upper sequent. 

\end{enumerate}

\end{definition}

By a ``reduced proof system'' relative to a goal $G$ let us
mean a calculus whose axioms are of the form
$\sequent{\Delta}{\top}$ and whose rules are the
RESTART, ATOMIC and BACKCHAIN rules relativized to $G$, \orr, \andr,
\impr\ and \somer. The main result of this section is then the
following:

\begin{theorem}\label{thm:reduced}
Let $\Gamma$ be a (multi)set of program clauses and let $G$ be a goal
under the syntax described for such formulas in this section. Then
$\sequent{\Gamma}{G}$ has a \Cbf-proof if and only if it has a proof
in the reduced proof system relative to $G$.
\end{theorem}

\begin{proof}
An immediate consequence of Theorem~\ref{thm:firstmodified} and
Lemmas~\ref{lem:secondmodified} and \ref{lem:backchain}.
\end{proof}

The reduced proof system provides the basic structure of the promised
procedure for constructing proofs for formulas in classical
logic. This 
procedure would simplify complex goals based on the rules 
\orr, \andr, \impr\ and \somer\ and would use an instance of the
RESTART, ATOMIC or BACKCHAIN rule on reaching an atomic formula. In a
practical rendition of this procedure, it will be necessary to delay
the choice of term to be used relative to the \somer\ rule. A
suitable delaying ability can, as usual, be obtained by using
a variable that can be later instantiated in conjunction with this
rule and by carrying out the instantiation by using unification in
the implementation of the ATOMIC and BACKCHAIN rules. 

The procedure described above can, of course, also be used to
find proofs for sequents of the form $\sequent{\Gamma}{G}$. An
interesting aspect of this procedure is that it reduces to others
described in the literature when (further) restrictions are placed on
the syntax of $G$ and the formulas in $\Gamma$. For example, suppose that
disjunction and the symbol $\bot$ are disallowed in the heads of
program clauses and implication is disallowed at the
top-level in goals. The logic being considered
reduces in this case to that of Horn clauses. From
Lemma~\ref{lem:sansorlimpr} and an 
examination of the proof of Lemma~\ref{lem:secondmodified}, it is
easily seen that the RESTART rule is redundant in this
context. Further, only the degenerate forms of the ATOMIC and
BACKCHAIN rules are relevant in this situation and that too in a form
where the possibility of $\bot$ being the head of a clause instance
need not be considered. Our procedure is equivalent under these
observations to the usual one employed for Horn clause logic. 
Along a different
direction, suppose the 
syntax of program clauses in the Horn clause setting is enriched by
permitting disjunctions in their heads, thereby producing the logic
underlying disjunctive logic programming \cite{NL95lics}. From an
examination of the proofs of Lemmas~\ref{lem:iproofexists} and
\ref{lem:secondmodified}, it becomes apparent that the RESTART rule is
redundant in this situation as well. The exclusion of this rule from
our proof procedure yields one that has the essential structure of the
Inheritance Near-Horn Prolog procedure (InH-Prolog) \cite{LR91rob,
  RLS91ilps}. Finally suppose 
that disjunction and $\bot$ is disallowed in the heads of program
clauses but that the syntax for these formulas and goals is otherwise
unaltered from the one presented at the beginning of this section. The
resulting goals subsume (conjunctions of) the N-clauses of
\cite{GaRe84}. In this context, the RESTART rule and only (restricted
versions of) the 
degenerate forms of the ATOMIC and BACKCHAIN rules are relevant and
our proof procedure reduces to (a simple generalization of) the
QNR-Prolog procedure described in \cite{Gab85}. In recent work
\cite{GR93}, Gabbay and Reyle have extended the QNR-Prolog procedure
to a fragment of classical logic that excludes only negative
occurrences of disjunctions. The preferred approach in \cite{GR93}
appears to be one that incorporates a run-time calculation of the
effects of static Herbrandization. The latter process eliminates 
essential universal quantifiers and the resulting fragment is thus 
contained in the one discussed in this paper. As indicated earlier,
the proof procedure presented in this section can be adapted in a
straightforward way to apply directly to this larger fragment and
would, in this form, subsume the mentioned one in
\cite{GR93}.

An important aspect of the proof procedure we have outlined above is
the directionality present in the backchaining rules used in it. We
note that the ability to impart this directionality to
these rules is also dependent of the augmentation of the assumption
set with the negation of the original goal formula. To see this, 
suppose that the formulas in the antecedents of the
sequents whose proofs we seek are either disjunctions of
atoms or of the form \[ (B_1 \land \ldots \land B_n) \supset (A_1 \lor
\ldots \lor A_m) \]
where the $B_i$s and $A_j$s are atomic and the succedents of these
sequents are conjunctions of atoms; despite the apparently severe
syntactic restrictions on the formulas, this context is of interest
because it corresponds to 
propositional classical logic under a translation to clausal
form. Now, it is easily seen that classical, intuitionistic and
uniform provability coincide for sequents of the described
kind. However, when a proof is sought for an atomic 
formula in this context, this formula does not always help in
determining what should be used from the assumption set. For example,
consider the 
sequent 
\[\sequent{p \lor q, p \supset r, q \supset r}{r}.\]
In constructing a \Ibf-proof for this sequent, the last rule that must
be used is an \orl\ that introduces the top-level connective in the
assumption formula $p\lor q$. At a deeper level, the inability to
use the (atomic) succedent formula to drive the proof search in this
situation arises from the fact that \orl\ rules may sometimes have to
appear after \impl\ rules in \Ibf-proofs. The augmentation of the
assumption set with the negation of the goal formula permits the \orl\
rule to be replaced by the \orlg\ rule, leading eventually to an
elimination of the mentioned dependency. An alternative approach,
which works within the original proof system, is to proceed as if 
\orl\ rules are not required in the proof being constructed and, when
they are determined to be actually necessary, to attempt to insert
them at an appropriate point in the proof. The modified problem
reduction format of 
Plaisted \cite{Plaisted88} is based on this approach and on noting
that the use of assumption formulas of the form $(B_1 \land \ldots
\land B_n) \supset A$ where $A$ is atomic can be driven, even in this
context, by the atomic formula for which proof is sought.

\section{Conclusion}\label{sec:conc}
We have examined the applicability of the notion of uniform provability to
classical logic in this paper. It is easily observed that this form of
derivation diverges from classical provability in the general
case. However, we have shown that if there are no essentially positive
occurrences of universal quantifiers in our formulas, then a modest,
sound, modification to the set of assumptions --- in particular, the
addition to them of the negation of the formula to be proved ---
results in a coincidence between uniform and 
classical provability. We have exploited this fact in outlining 
a proof procedure for classical logic. The procedure that we have
described subsumes several previously proposed ones for different
subsets of classical logic. The uniform proof notion appears, in this
sense, to be a unifying principle behind proof search in this logical
setting. 

The discussions in this paper suggest other directions for
further investigation. At a pragmatic level, it is of interest to
develop, and to experiment with, an actual proof procedure based on the
ideas presented here. Another matter worthy of consideration is the
usefulness of the uniform proof notion and the general approach
described here in structuring proof search in intuitionistic logic.

\section{Acknowledgements}\label{sec:ack}

This work has grown out of a prior collaboration with Donald
Loveland \cite{LN94,NL95lics} and has been enriched by his
suggestions. We are also grateful
to Robert St\"{a}rk whose comments helped us discover
an error in an earlier version of this paper. This work was started
when the 
author was visiting Ludwig-Maximilians-Universit\"{a}t
M\"{u}nchen. Support from the Konrad Zuse-Programm administered by the
  Deutsche Akademischer Austauschdienst during this visit and
  subsequently from NSF Grant CCR-92-08465 is gratefully acknowledged.

       \end{document}